\documentclass[twocolumn,pra,showpacs]{revtex4}

\usepackage{graphicx}
\usepackage{dcolumn}
\usepackage{bm}
\usepackage{nicefrac}
\usepackage{amsmath}

\newcommand{\eref}[1]{(\ref{#1})}
\newcommand{\Eref}[1]{Eq.~(\ref{#1})}

\newcommand{\rtw}{\rightarrow}
\newcommand{\cm}{cm$^{-1}$}
\newcommand{\vect}[1]{{\bm #1}}

\newcommand{\half}{{\nicefrac{1}{2}}}
\newcommand{\threehalf}{{\nicefrac{3}{2}}}
\newcommand{\photon}{\omega}
\newcommand{\hfs}{{h\!f\!s}}
\newcommand{\fs}{{f\!s}}
\newcommand{\omegahfs}{\omega_{h\!f\!s}}
\newcommand{\omegafs}{\omega_{f\!s}}

\newcommand{\sixj}[6]{\left\{ \begin{array}{ccc} #1 & #2 & #3 \\ #4 & #5 & #6 \end{array} \right\} }

\begin{document}

\title{Magnetic blackbody shift of hyperfine transitions for atomic clocks}

\author{J. C. Berengut}
\affiliation{School of Physics, University of New South Wales, Sydney 2052, Australia}
\author{V. V. Flambaum}
\affiliation{School of Physics, University of New South Wales, Sydney 2052, Australia}
\author{J. King-Lacroix}
\affiliation{School of Physics, University of New South Wales, Sydney 2052, Australia}

\date{12 October 2009}

\begin{abstract}

We derive an expression for the magnetic blackbody shift of hyperfine transitions such as the cesium primary reference transition which defines the second. The shift is found to be a complicated function of temperature, and has a $T^2$ dependence only in the high-temperature limit. We also calculate the shift of ground-state $p_\half$ hyperfine transitions which have been proposed as new atomic clock transitions. In this case interaction with the $p_\threehalf$ fine-structure multiplet may be the dominant effect.

\end{abstract}


\maketitle

The frequency of the ground-state hyperfine transitions used in atomic clocks (such as the cesium primary standard) are known to be temperature dependent~\cite{itano82pra}. For this reason the SI second is defined at $0\,$K, and at any finite temperature the blackbody shift must be taken into account. Temperature fluctuation of the laboratory is a major portion of the clock error budget~\cite{jefferts07appa}, therefore the NIST-F2 cesium fountain, currently under construction, will be cooled to $77\,$K to reduce the blackbody shift.

Recently there was some disagreement in the literature over the size of the electric blackbody radiation shift in cesium. Early measurements and \emph{ab initio} calculations support a value about 10\% higher than later measurements and semiempirical calculations (see \cite{angstmann06pra} for references). On the theory side, this seems to have been resolved~\cite{angstmann06prl,beloy06prl,angstmann06pra} in favour of the larger values. As the temperature of the experiment is reduced in the future the magnetic blackbody shift ($\sim T^2$) will become more important relative to the electric shift ($\sim T^4$). Hence this reassessment of the magnetic blackbody shift.

In this paper we present a derivation of the magnetic blackbody shift of ground-state hyperfine transitions that is valid at all temperatures (not just in the high-temperature limit). We calculate the effect for $s_\half$ hyperfine transitions such as the $6s_\half$ ($F=3\rightarrow 4$) $^{133}$Cs transition which defines the second (there are many other such clocks, including $^{87}$Rb, $^{171}$Yb$^+$, and $^{199}$Hg$^+$). We find that the simple scaling law of the blackbody shift $\Delta\omegahfs\sim T^2$ is only valid at high temperatures. Additionally we calculate the shift for $p_\half$ hyperfine transitions which have been proposed as clock references~\cite{beloy09prl}. We show that interaction with the $p_\threehalf$ fine-structure multiplet must be considered.

The magnitude of the magnetic blackbody field is (atomic units $\hbar = e = m_e = 1$)
\begin{equation}
\label{eq:bbspectrum}
B^2 (\photon)\, d\photon = \frac{8 \alpha^3}{\pi} \frac{\photon^3 d\photon}{e^{\photon/kT} - 1}
\end{equation}
An oscillating magnetic field $\vect{B}(\photon)\cos(\photon t)$ affects an atomic energy level via the time dependent perturbation
\begin{equation}
V(\photon, t) = -\vect{\mu}\cdot\vect{B}(\photon)\cos(\photon t)
\end{equation}
where $\vect{\mu}$ is the magnetic dipole moment of the system. The energy is affected in the second order of perturbation theory (see, e.g.~\cite{manakov86prep})
\begin{equation}
\label{eq:deltaE}
\Delta E_a = \frac{1}{4} \sum_n \left(
	\frac{\left| \left< a |\hat v^+| n \right>\right|^2}{E_a - E_n + \photon} +
	\frac{\left| \left< a |\hat v^-| n \right>\right|^2}{E_a - E_n - \photon}
	\right)
\end{equation}
where $\hat v^+ = \hat v^- =-\vect{\mu}\cdot\vect{B}(\photon)$. For an atom with a single electron above closed shells \mbox{$\vect{\mu} = -\mu_B(\vect{L} + g_s\vect{S})$} with $g_s = 2$ and $\mu_B = \alpha/2$ in atomic units. A general expression for this case is presented in the Appendix.

We first examine the case of a single $s_\half$ orbital split by the hyperfine interaction with a nuclear spin $I$. In this case there are only two levels of interest, with $F = I + \half$ and $F = I - \half$; the next level will be separated by several orders-of-magnitude more than the splitting (in the $^{133}$Cs case by a factor of $10^5$). Then $\vect{\mu} = -\mu_B g_s \vect{S}$, and one obtains
\begin{eqnarray}
\Delta E_{I+\half} &=& \left( \frac{g_s \mu_B}{2} \right)^2
	\frac{I}{2I+1}\frac{B^2}{3}
	\frac{2\omegahfs}{\omegahfs^2 - \photon^2} \\
\Delta E_{I-\half} &=& \left( \frac{g_s \mu_B}{2} \right)^2
	\frac{I+1}{2I+1}\frac{B^2}{3}
	\frac{-2\omegahfs}{\omegahfs^2 - \photon^2}
\end{eqnarray}
where $\omegahfs = E_{I+\half} - E_{I-\half}$ is the hyperfine splitting of the $s$-state. The total blackbody shift is obtained by integrating this shift over the blackbody spectrum~\eref{eq:bbspectrum}:
\begin{eqnarray}
\frac{\Delta \omegahfs}{\omegahfs}
	&=& \frac{\alpha^2}{6} \int \frac{B^2(\photon)}{\omegahfs^2 - \photon^2} d\photon \\
	&=& -\frac{4 \alpha^5}{3\pi}(kT)^2 \int\frac{x^3\,dx}{(e^x-1)(x^2 - a^2)}
	\label{eq:integral}
\end{eqnarray}
where $a \equiv \omegahfs/kT$. Note that the fractional blackbody shift does not depend on the nuclear spin $I$.

\begin{figure}[tb]
\includegraphics{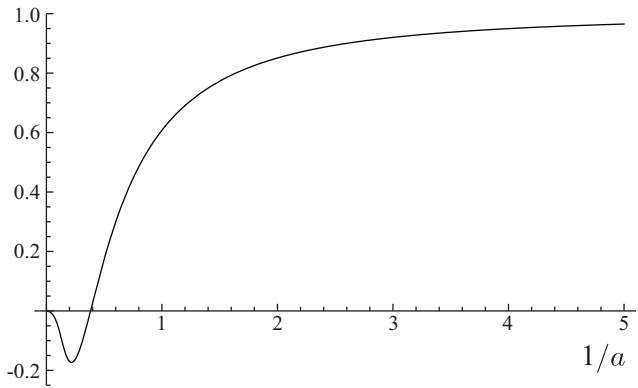}
\caption{\label{fig:integral} The integral of \Eref{eq:chi}, normalised to the high-temperature limit:
$\frac{6}{\pi^2}\chi(a)=\frac{6}{\pi^2}\int\frac{x^3\,dx}{(e^x-1)(x^2 - a^2)}$\ . Note that the abscissa is $1/a$: temperature increases to the right.
}
\end{figure}

At high temperatures, $a \ll 1$, the integral in \eref{eq:integral} is analytic and one obtains
\[
\frac{\Delta \omegahfs}{\omegahfs} = -\frac{2\pi}{9} \alpha^5 (kT)^2, \quad kT \gg \omegahfs
\]
in agreement with Ref.~\cite{itano82pra}. By contrast, in the low temperature limit the blackbody shift has a $(kT)^4$ dependence (and is of opposite sign):
\[
\frac{\Delta \omegahfs}{\omegahfs} = \frac{4 \pi^3}{45} \alpha^5 \frac{(kT)^4}{\omegahfs^2}, \quad kT \ll \omegahfs .
\]
Of course, in general one can simply calculate the shift numerically; this can be done more easily by subtracting the pole. Defining $F(x)=x^3/(e^x -1)$ the integral can be rewritten
\begin{equation}
\label{eq:chi}
\chi(a)=\int_0^\infty \frac{F(x) - F(a)}{x^2 - a^2} dx
\end{equation}
since the contribution of the second term $F(a)/(x^2-a^2)$ is zero. $\chi(a)$ is continuous and differentiable everywhere on the positive real axis.
A graph of the integral is presented in Fig.~\ref{fig:integral}. Using \Eref{eq:integral} we obtain, for $^{133}$Cs at $300\,$K, $\Delta\omegahfs/\omegahfs = -1.304\times 10^{-17}$.

We now turn our attention to other clocks that use the hyperfine splitting of a ground $p_\half$-wave state as the reference frequency, such as those proposed in~\cite{beloy09prl}. In this case the blackbody radiation will again cause attraction (or repulsion) between the two $p_\half$ levels, in a similar fashion to the $s_\half$-wave case. Using the results of the Appendix, one obtains the first term in \Eref{eq:pshift}.

However in the $p_\half$ case there will be an additional shift due to interaction with the $p_\threehalf$ level. In fact, there is the possibility of some ``enhancement'' of the blackbody effect here as can be seen from \Eref{eq:deltaE}: in the case where $\photon \gtrsim E_a - E_n = \omegafs$ the shift will be of order $\Delta E_a \sim \omegafs/\photon^2$, which can be larger than the shift due to mixing of the $p_\half$ hyperfine states by a factor $\omegafs/\omegahfs$. However one finds that the shift for both the $F=I+\half$ and $F=I-\half$ levels due to the $p_\threehalf$ levels is identical in second order. To go beyond second order we have included the differences in the energy denominators between different hyperfine components. This affects the blackbody shifts at the level $\omegahfs/\omegafs$, which cancels the enhancement mentioned earlier. One obtains for the interaction
\begin{align}
\label{eq:pshift}
\frac{\Delta \omegahfs}{\omegahfs}
	&= -\frac{4\alpha^5}{27\pi} (kT)^2 \chi(a_\hfs) \\
	&\quad -\frac{2\alpha^2}{9\pi} (kT)^2
	 \left( 1 - \frac{5(2I+1)}{6}\frac{A}{\omegahfs} \right) \nonumber \\
	&\qquad\cdot \bigl( \chi(a_\fs) + a_\fs\, \chi'(a_\fs) \bigr) \nonumber
\end{align}
where $a_\hfs = \omegahfs/kT$ and $a_\fs = \omegafs/kT$. The second term, proportional to $\chi(a_\fs) + a_\fs \chi'(a_\fs)$, shows the effect of the $p_\threehalf$ levels on $\omegahfs$.
The first part of the second term is due to the energy difference of the $p_\half$ levels in the energy denominator, while the second part ($\sim A/\omegahfs$) is due to to splitting of the $p_\threehalf$ levels. Here $A$ is the magnetic-dipole hyperfine constant of the $p_\threehalf$ levels. Note that the electric-quadrupole terms $B$ cancel.

\begin{figure}[tb]
\includegraphics{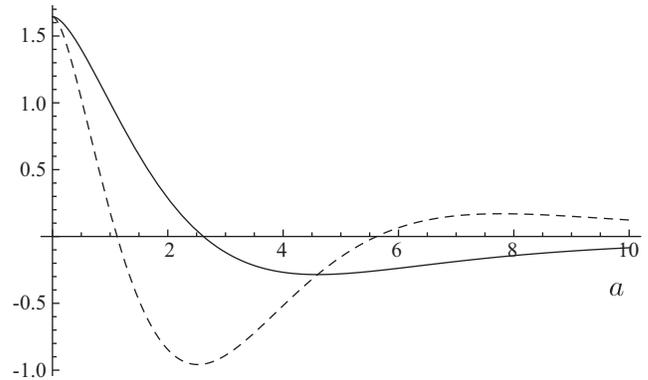}
\caption{\label{fig:bothchis} Solid line: $\chi(a)$ from \Eref{eq:chi}; dashed line $\chi(a) + a \chi'(a)$ (used in \Eref{eq:pshift}). The high-temperature limit $a\rtw 0$ of both graphs is $\pi^2/6$.
}
\end{figure}

The function $\chi'(a)$ in \eref{eq:pshift} arises from the expansion of the energy denominators and is defined by
\begin{equation}
\chi'(a_\fs) = \frac{\chi(a_\fs + a_\hfs) - \chi(a_\fs)}{a_\hfs}
\end{equation}
in the limit $a_\hfs/a_\fs = \omegahfs/\omegafs \rtw 0$. We present a graph of 
$\chi(a)+a \chi'(a)$ in Fig.~\ref{fig:bothchis}.

In the high temperature limit  $kT \gg \omegafs$, $\chi(a \rtw 0) = \pi^2/6$ and $\chi'(a \rtw 0) = 0$ so from \Eref{eq:pshift} we obtain
\begin{align}
\label{eq:pshiftlimit}
\frac{\Delta \omegahfs}{\omegahfs}
 &= -\frac{\pi}{81}\, \alpha^5 (kT)^2 \left( 2 + 3 - \frac{5(2I+1)}{2}\frac{A}{\omegahfs}\right) \nonumber \\
 &= -\frac{5\pi}{81}\, \alpha^5 (kT)^2 \left( 1 - \frac{2I+1}{2}\frac{A}{\omegahfs}\right)
\end{align}
Equations~(\ref{eq:pshift},\ref{eq:pshiftlimit}) show that the $p_\threehalf$ levels must be included when calculating the magnetic blackbody shift unless $kT \ll \omegafs$: depending on the system considered (i.e. the value of $I$ and $A/\omegahfs$) it may be the dominant effect. In the case of Al, $A/\omegahfs = 0.063$ \cite{lew49pr,lew53pr} and $I=5/2$, therefore the last term in \eref{eq:pshiftlimit} is approximately 0.19. However in Al, $\omegafs = 112\,\text{\cm} = 162\,$K, therefore at $300\,$K, $a_\fs = 0.54$ and $\chi(a) + a \chi'(a) \approx 0.96 \approx 0.59\, \pi^2/6$, so clearly the high-temperature limit is not appropriate. At $300\,$K one obtains $\Delta\omegahfs/\omegahfs = -2.32\times 10^{-18}$.

Our treatment of the interaction with the $p_\threehalf$ levels takes into account only the largest terms in third-order of perturbation theory (second-order in $V$ and first-order in the hyperfine interaction); numerical calculation of off-diagonal hyperfine interaction constants is beyond the scope of this work. In any case usually the non-diagonal hyperfine matrix element $\left<p_\half \left| H_{h\!f\!s} \right| p_\threehalf\right>$ is significantly smaller than the diagonal one ($\omegahfs$ for $p_\half$), therefore we do not expect the result to change significantly. However when a clock is produced a more accurate third-order calculation will be necessary. In the meantime the last term of \eref{eq:pshift} ($\sim A/\omegahfs$) may be considered an estimate of the error.

We thank W. Itano for useful discussions.
This work is supported by the Australian Research Council.

\appendix*
\section{Magnetic blackbody shift}

The total magnetic blackbody shift of a level $\left| a\,m_a \right>$ is given by equation~\eref{eq:deltaE}. For electrons $\vect{\mu} = -\mu_B (\vect{J} + \vect{S})$ if we neglect the anomalous magnetic moment (i.e. $g_s=2$). Then
\begin{align}
\begin{split}
\Delta E_a &= \frac{1}{4}\sum_{b\,m_b}
 \left|\left< b\,m_b\left| \mu_B (\vect{J} + \vect{S})\cdot\vect{B} \right| a\,m_a\right>\right|^2 \\
 &\qquad\qquad \cdot \left( \frac{1}{E_a-E_b+\photon} + \frac{1}{E_a-E_b-\photon} \right) \nonumber
\end{split}\\
 &= \frac{\mu_B^2 B^2(\photon)}{6} \sum_b C_{ba}\, \frac{E_a - E_b}{(E_a-E_b)^2 - \photon^2}
\end{align}
Consider the interaction of levels in an atom with nuclear spin $I$.
If we denote the angular quantum numbers of $\left| a \right>$ with $L$, $J$ and $F$, and those of $\left| b \right>$ with $L'$, $J'$ and $F'$, then
\begin{widetext} 
\begin{equation}
C_{ba} = [F'] \sixj{F'}{F}{1}{J}{J'}{I}^2
	\left( \delta_{J,J'}\sqrt{J(J+1)(2J+1)}\ +
   \delta_{L,L'}(-1)^{P}[J,J']^{\half}\sixj{J'}{J}{1}{\half}{\half}{L}
\sqrt{\frac{3}{2}} \right)^2
\end{equation}
where $P=J+L+\half+F+F'+2I+1$ and the square brackets $[J]=(2J+1)$.
\end{widetext}

\bibliography{references}

\end{document}